\documentclass[12pt]{article}

\usepackage{scicite}
\usepackage{times}

\usepackage{amsmath,amsfonts,amsthm}
\PassOptionsToPackage{hyphens}{url}\usepackage[hidelinks]{hyperref}
\usepackage[detect-all]{siunitx}
\usepackage{graphicx}
\graphicspath{ {Figures/} }
\usepackage{subcaption}
\usepackage{eufrak}
\usepackage{pdflscape}
\usepackage{fancyhdr}
\usepackage{xcolor}
\newenvironment{sciabstract}{%
\begin{quote} \bf}
{\end{quote}}

\title{Distributed Sensing Along Fibres for Smart Clothing}

\author
{Brett C. Hannigan$^{1}$, Tyler J. Cuthbert$^{1}$, Chakaveh Ahmadizadeh$^{1}$, Carlo Menon$^{1\ast}$\\
\\
\normalsize{$^{1}$Department of Health Sciences and Technology, ETH Zürich,}\\
\normalsize{Lengghalde 5, 8008 Zürich, Switzerland}\\
\normalsize{$^\ast$To whom correspondence should be addressed; E-mail: carlo.menon@hest.ethz.ch.}
}

\date{}

\begin{document} 


\baselineskip24pt


\maketitle 


\begin{sciabstract}
	Textile sensors transform our everyday clothing into a means to track movement and bio-signals in a completely unobtrusive way. One major hindrance to the adoption of ``smart'' clothing is the difficulty encountered with connections and space when scaling up the number of sensors. There is a lack of research addressing a key limitation in wearable electronics: connections between rigid and textile elements are often unreliable and they require interfacing sensors in a way incompatible with textile mass production methods. We introduce a prototype garment, compact readout circuit, and algorithm to measure localized strain along multiple regions of a fibre. We employ a helical auxetic yarn sensor with tunable sensitivity along its length to selectively respond to strain signals. We demonstrate distributed sensing in clothing, monitoring arm joint angles from a single continuous fibre. Compared to optical motion capture, we achieve around \SI{5}{\degree} error in reconstructing shoulder, elbow, and wrist joint angles. 
\end{sciabstract}

\section{Introduction}\label{intro}

Soft strain sensors have become an active research area in part because their use in ``smart textiles'' is needed to achieve movement tracking in next generation wearable devices. Topical applications of wearable strain sensors include kinematics monitoring, pose estimation, and the measurement of mechanically-transduced biosignals (e.g. heart rate, speech) \cite{Kim2019,He2019,Yin2019,Madhavan2022}. Soft sensor arrays have been used to map pressure, to provide feedback for soft robotics, and in human interface devices \cite{Zhu2022,Alian2023}. Sensing movement through textiles is unique and has potential for high impact in advancing our ability to track athletics and monitor health \cite{Boesel2020}. Clothing---in comparison to other wearable technology that is attached directly to the body or uses rigid components such as exoskeletons---possesses inherent variability including folds, non-homogeneity in body contact (e.g., tightness, looseness), and confounding mechanical strain (e.g., pressure, stretch, twisting). To combat this limitation of clothing, there has been a focus on increasing the number of sensing elements around the location of focus to increase monitoring accuracy. For example, the current state-of-the-art in upper-body joint angle tracking with wearable sensors requires 3--8 sensors per joint for angle estimation \cite{Esfahani2018,Jin2020}. To have high spatial resolution without an impractically high number of connections, sensor arrays are often used. Ever-denser arrays are pursued for finer resolution in soft pressure sensors \cite{Kim2020}. However, as the number of sensing locations is increased, complexity increases because many sensors and connections are required. The factors limiting the scaling up of the density of sensors are often the amount of space available on the garment for sensors and interconnections and the method for reliably connecting these sensors and accessing their signals. Connections between textiles and rigid electronics are particularly troublesome because the high stress at the interface makes them prone to failure \cite{Niu2019,Castano2014}. Arrays introduce other problems such as crosstalk between channels, which can limit their resolution \cite{Saxena2011}. 

What makes textile sensors most attractive is their ability to transform the clothing we wear every day into functionalized devices with absolutely minimal obtrusiveness. Full textile integration is necessary to achieve widespread adoption of ubiquitous ``smart'' garments in sports, rehabilitation, occupational health, and everyday life. Such integration requires the reduction and eventual elimination of rigid electronic components on the garment and constrains the design to what would be attainable with entrenched industrial textile production methods. The sustainability of functional textiles is subject to increased recent scrutiny because of the large amount of waste associated with their eventual disposal coupled with the various advanced materials e-textiles may contain \cite{Dulal2022}. Separation of these materials for recycling is difficult, so there exists two main ways to address this challenge: making separation at end-of-life easier or moving toward using a single material \cite{Kohler2011}. The distributed sensing approach has promise to address some of these problems by replacing discrete interconnects and electronics in the garment with single fibre sensors. In this case, the sensing fibres themselves should ideally consist of the same materials as the bulk garment, preferentially using biodegradable or easily separable polymers, and be made with cleaner fabrication processes. Continuous fibres able to sense at multiple points along their length that can be woven or knit into fabric would greatly reduce connection issues, allow greater sensor density, potentially increase sustainability, and maintain compatibility with established textile processes. It follows that achieving these goals would thereby increase the reliability and lower the cost of ``smart'' textile garments.

\paragraph{Distributed Strain Sensing} Distributed sensing is a promising solution to these aforementioned problems encountered when scaling up strain and pressure sensing garments. A distributed sensing system permits multiple measurements out of a single sensor element, each localized in space. Unlike approaches using multiple sensors connected by wires, multiplexers, or switches, the electrical connectivity is greatly simplified \cite{Nilsson2000,Tairych2017}. Distributed sensing can also be used to increase sensitivity \cite{Nesser2021,Nesser2022}, allow arbitrarily high spatial resolution \cite{Sonar2018}, and simplify 2-D pressure or strain sensitive arrays \cite{Nilsson2000,Sonar2018,Xu2015}. Capacitive strain sensors have garnered recent interest because the configuration is possible in yarn, coaxial, or twisted fibre form and possesses linear response to strain, which is ideal for integration into textiles \cite{Lee2021,Yu2019,Cooper2017,Zhang2019d}. Distributed strain sensing has been shown with promising bench tests using chains of soft capacitive strain sensors \cite{Tairych2017}. Segmented soft capacitive sensors can be used to isolate strain at individual segments through microfracturing of piezoresistive electrodes \cite{Nesser2022}. Distributed pressure sensitive arrays have been demonstrated using frequency domain measurements of RLC resonator chains with a breadth of sensing applications \cite{Kim2022a}. The RLC approach has yet to address the elimination of rigid electronic components on the sensing element and it requires a parallel connectivity of multiple resonant circuits, meaning that it would be difficult to both implement using fully soft electronics and transform into a linear fibre. Another method we may classify as `pseudo-distributed' sensing uses algorithms to extrapolate localized strain from the DC resistance of a piezoresistive sensor \cite{Kim2018a}. This technique shows impressive results for classifying sets of repeatable patterns using a piezoresistive nanomesh sensor \cite{Kim2022} but it is fundamentally not possible to reconstruct arbitrary, independent localized strains from single-port DC resistance measurements. However, piezoresistive strain sensors (e.g. using graphene nanofilms) \cite{Cao2021, Li2021a} may also be used for distributed sensing with frequency-domain impedance measurements. Applying techniques from electrical impedance tomography, a sensor mesh made from kinesiology tape coated with graphene nanofilms can be interrogated to obtain a continuous 2-D strain map \cite{Lin2021}. This greatly increases the information collected from the sensing mesh but requires a series of perimeter connections. Here, we aim instead to eliminate the majority of unreliable connection points and increase textile integration with a 1-D fibre. Distributed sensing is yet to be integrated as a fibre into a textile garment or as a wearable system. We desired to build upon these results and apply distributed strain sensing methods along a fibre that is highly suitable for textile integration.

\begin{figure}
	\centering
	\includegraphics[width=88.9mm]{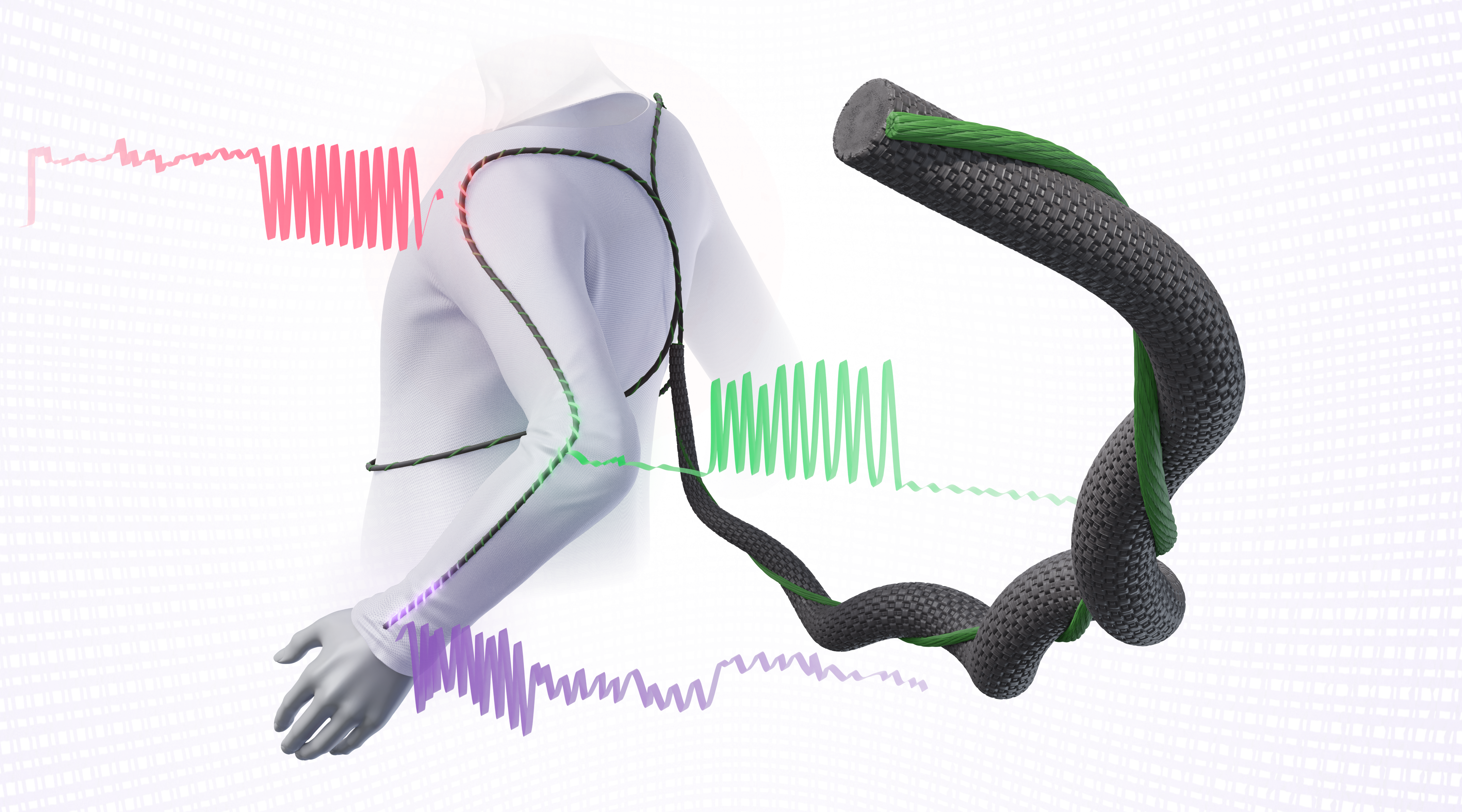}
	\caption{A conceptual illustration of our distributed sensing approach used to measure the three major joint angles of the arm, with an envisioned extension of the sensing fibre around the torso.} \label{fig:conceptual-graphic}
\end{figure}

In this paper, we demonstrate a system to enable distributed sensing along a strain sensor and a prototype garment that monitors the three major arm joint angles with one continuous sensing fibre. Our solution isolates the single pair of connections to one end of the stretchable fibre, so that the fibre may be sewn or woven into a textile without connections or wires in the fabric. This fibre sensor design allows the single remaining connection to be centralized at one end in a proximal hub location. It compares favourably to some other fibre strain sensor designs that require wires from both ends or access to an inner layer (e.g. coaxial capacitive fibres), which brings much greater complexity to textile manufacturing. We can envision extending the prototype to include many functionalized fibres in the garment, for example to greatly increase sensor density or instrument multiple extremities, as illustrated conceptually in Fig.~\ref{fig:conceptual-graphic}. We further developed a compact electronic impedance analyzer circuit to collect high-speed impedance measurements at multiple frequencies in parallel and use this device to develop an algorithm and test fixture to quantify strain reconstruction performance more rigorously. We show, for the first time, the application of a single-fibre capacitive sensor with tunable sensitivity at different locations along its length, which further increases the capabilities of our distributed sensing system. We combine these into a prototype garment that can monitor the angles of the shoulder, elbow, and wrist joints with high accuracy on a single channel.

\section{Results} \label{sec:results}

\begin{figure}
	\centering
	\includegraphics[width=\textwidth]{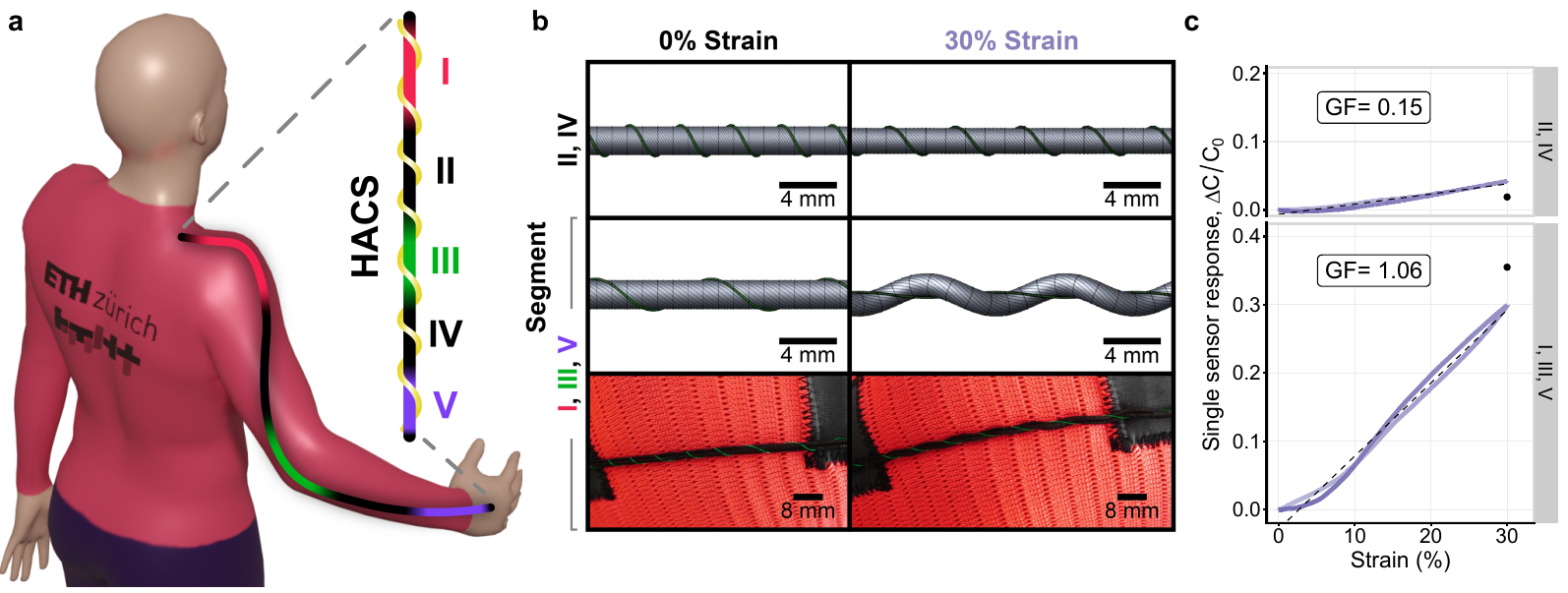}
	\caption{\textbf{Prototype garment and sensing fibre.} \textbf{a}, Rendering of the path of the sensor along the arm, where I, III, and V indicate the high sensitivity regions around the shoulder, elbow, and wrist joints, respectively. \textbf{b}, Top: geometric modelling of the HACS sensor low-sensitivity (II \& IV: \SI{3.5}{\milli\metre} pitch) and high sensitivity (I, III, \& V: \SI{8}{\milli\metre} pitch) regions under relaxed and 30\% strain conditions; Bottom: photographs of the unstrained and strained sensor in the garment at the elbow joint. \textbf{c}, Response versus strain plots (solid), gauge factor (dashed black), and computational model predictions (round black marker) for the respective HACS sensitivity regions (model adapted from our previous work \cite{Cuthbert2023}).} \label{fig:garment-regions}
\end{figure}

\paragraph{A Sensing Fibre with Selective Response to Strain} %\label{sec:fibre-garment}
While in the general case, distributed sensing methods may be applied to various transduced phenomena of interest---potentially with a multi-dimensional sensor configuration---here we limit our focus to strain sensing distributed in 1-D along the axis of a fibre, as these are the atomic components of textiles. A distributed strain sensing fibre designed to be incorporated into textiles should be highly sensitive and compatible with existing textile processing methods. While the fibre itself may follow a long path along the garment, the desired strain information to isolate a movement of interest is often concentrated in few areas along the fibre's length (e.g. around the joints). Thus, we desired our sensing fibre to have spatial specificity to strain at certain locations. We previously reported a new capacitive fibre strain sensing modality \cite{Cuthbert2023} having some attractive features that can be utilized for distributed sensing applications. These are: (i) leveraging the auxetic behaviour of a helical yarn complex to achieve a higher sensitivity than expected for typical capacitive strain sensors; (ii) the ability to tune this sensitivity by manipulating the helical pitch; and (iii) high robustness to stress and suitability for a reel-to-reel manufacturing process. These helical auxetic capacitive sensor (HACS) fibres are composed of an inextensible conductor coiled around a conductive elastomer core. The core is an ordinary polyester wrapped elastane fibre, coated with polypyrrole using a low-waste vapour phase process. Copper wire used as the inextensible conductor coil, as it possesses the ideal mechanical properties as well as an insulating enamel layer to achieve the desired capacitor configuration. The highly robust copper wire wrapping also limits the maximum strain of the sensor fibre, so that the polymer core does not plastically deform. Exchanging the copper wire component for a conductive yarn/thread is possible, although not within the scope of this research. Upon straining, the elastomer core stretches, while the inextensible wrapping increases in pitch and decreases in helical diameter. At higher strains, the inextensible component becomes nearly straight, causing the coils to ``flip'' so that the elastomer is coiled around it. The ``flipping'' increases the overall outer diameter with increasing strain; this is an example of auxetic material behaviour. These mechanics are shown in Fig.~\ref{fig:garment-regions}b and can allow one to access a larger gauge factor (GF) than ordinarily seen for capacitive strain sensors, which are typically limited by geometry to unity GF \cite{Amjadi2016}. The GF is defined in equation~(\ref{eq:gf}), where $C(\varepsilon)$ is the capacitance at strain $\varepsilon=\ell/\ell_0$ and $C(0)$ is the unstrained capacitance. Full details about the sensor fibre fabrication may be found in the Materials and Methods section. More important for this work than sheer sensitivity is the ability to manipulate the gauge factor along the sensor by adjusting the helical pitch. Our previous HACS were each only of a single pitch and therefore a single sensitivity. In this research, we modified the HACS such that we changed the pitch throughout a single fibre to achieve different sensitivities in different locations. This is a novel approach to fabricating HACS and employing our previous research. 

\begin{equation}
GF = \frac{\left(C(\varepsilon)-C(0)\right)/C(0)}{\varepsilon} \label{eq:gf}
\end{equation}

We devised a prototype garment by instrumenting a long-sleeved athletic shirt with a HACS fibre running from the shoulder to wrist (approximate path shown in Fig.~\ref{fig:garment-regions}a). We desire to monitor the three major angles of the arm as a proof of concept: shoulder adduction, elbow flexion, and wrist flexion. As described in \cite{Cuthbert2023}, the GF of these fibre sensors may be tuned by adjusting the helical pitch. We decided to use this to our advantage by manipulating the pitch in space along a single continuous fibre and more selectively respond to strains only at the joints. We chose a longer, \SI{8}{\milli\metre} pitch enabling high sensitivity for the regions of interest---three \SI{8}{\centi\metre} segments around the shoulder, elbow, and wrist joints (labelled as I, III, and V in Fig.~\ref{fig:garment-regions}). Because the maximum length of the sensing fibre is governed by the straightened length of the inextensible wrapping, which decreases with longer pitch values, there is a trade-off between attainable sensitivity and maximum strain. We chose the \SI{8}{\milli\metre} pitch value to maximize the GF while still allowing a sufficient strain range that covers what is expected in textiles. For the rest of the sensor, a shorter \SI{3.5}{\milli\metre} pitch ensures maximum insensitivity to strain at undesired locations (labelled as II and IV in Fig.~\ref{fig:garment-regions}). This type of sensor fibre has been shown to be relatively insensitive to torsion, with less than 20\% change in capacitance during a full 360$^\circ$ of torsion \cite{Cuthbert2023}---torsion near this amount would not be possible when affixed to the garment. We  investigated the effect of simultaneous strain and bending across different physiologically plausible bend radii in order to evaluate the fibre sensor's response to bending/stretching like that seen across the joints. The results show a very slight increase in signal for small bend radii, but the variability between samples is generally higher than that attributable to bending (see Section~S4 and Fig.~S7 in the Supplementary Materials). We further performed controlled tests to examine any degradation of sensitivity attributable to wear during cyclic durability tests. We observed very stable response and gauge factor before and after 1~000 cycles of straining to 20\% (see Section~S5 and Fig.~S8 in the Supplementary Materials). The geometry of these sensor regions when relaxed and subjected to strain are rendered using the geometrical model from \cite{Cuthbert2023} and are compared in Fig.~\ref{fig:garment-regions}b. The model predicts GFs of 1.18 and 0.06 for the highly-sensitive and insensitive regions, respectively. Universal testing machine (UTM) tests closely agreed with respective GFs of 1.06 and 0.15, shown in Fig.~\ref{fig:garment-regions}c. Next, we briefly describe the principle by which we may discriminate localized strain along these sensing fibres using impedance measurements.

\paragraph{Distributed Capacitive Strain Sensing} %\label{sec:distributed-capacitive}
To use the HACS fibre in a distributed sensing application, we need to identify the mechanism by which strain at various segments along the fibre may be differentiated. As mentioned previously, distributed sensing along a resistance-capacitance (RC) transmission line model has been demonstrated on the bench \cite{Tairych2018a}. In the following section, we will characterize short, single sensors to determine their response to strain. Using these results, we model the predicted behaviour of a chain of these sensors, using it to choose an excitation frequency band that well differentiates the localization of strain to each segment. We verify the model with measurements from the UTM and an inductance-capacitance-resistance (LCR) meter. Finally, we validate our readout electronics against this reference.

\begin{figure*}
	\centering
	\makebox[\textwidth][c]{\includegraphics[width=180mm]{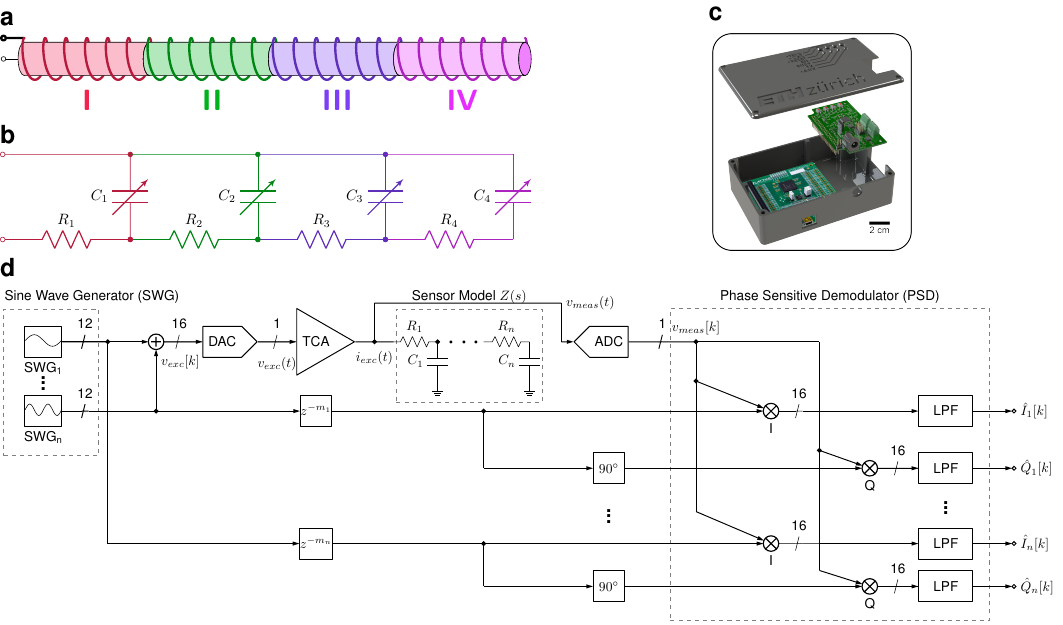}}
	\caption{\textbf{Discretized sensor schematic and electronics overview and function.} \textbf{a}, Schematic of sensor connectivity, split into four sensing regions. \textbf{b}, The corresponding RC ladder model circuit. \textbf{c}, Exploded view of the readout electronics, including enclosure, analog front-end printed circuit board, and field-programmable gate array (FPGA) board. \textbf{d}, Block diagram overview of the entire signal processing pathway with $n$ frequency channels, including sine wave generator (SWG), digital-to-analog converter (DAC), sensor model circuit, analog-to-digital converter (ADC), phase-sensitive demodulator (PSD), and filtering blocks.} \label{fig:system}
\end{figure*}

Like many sensors that employ capacitance to transduce strain, our fibre sensors can be considered as an electrical transmission line. The capacitance and resistance along a length of fibre is distributed in infinitesimally small segments. By approximating these with a finite number of lumped elements, we can probe a sensor with varying frequencies and use the response to estimate capacitance---and thus localized strain---at certain regions along its length. A schematic of a HACS discretized into $n=4$ segments and the corresponding transmission RC line model circuit are shown in Fig.~\ref{fig:system}a and Fig.~\ref{fig:system}b, respectively. An expression for the input impedance $Z_i$ at stage $i \in \{ 1, \ldots, n \}$ of an RC transmission line with this structure may be generated recursively as follows:

\begin{equation} \label{eq:transmission-line}
	Z_i(\omega) = 
	\begin{cases}
		\frac{1}{j\omega C_i + \frac{1}{Z_{i+1}(\omega)}} + R_i &1 \leq i < n\\
		\frac{1}{j\omega C_{n}} + R_{n} &i = n
	\end{cases}
\end{equation}

\begin{figure}
	\centering
	\makebox[\textwidth][c]{\includegraphics[width=180mm]{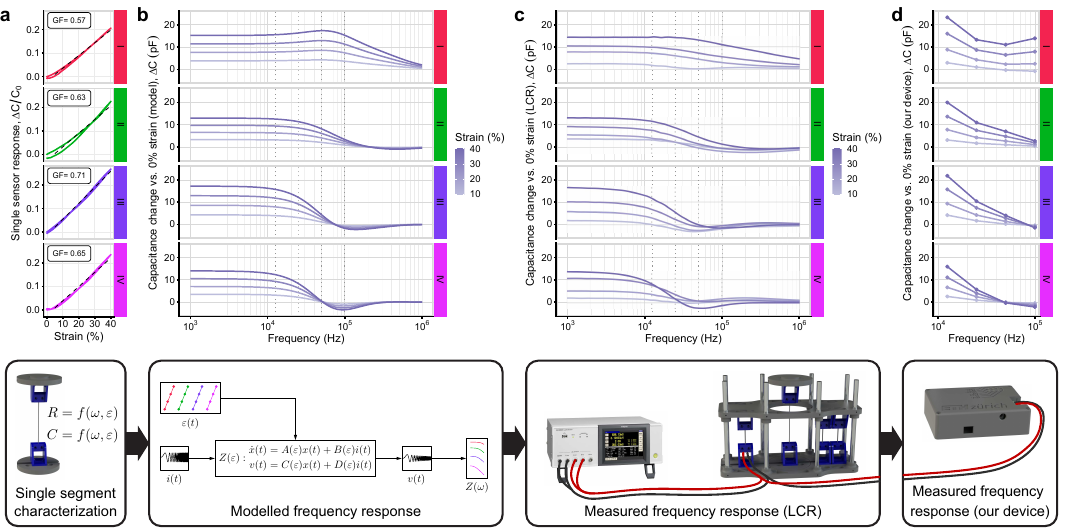}}
	\caption{\textbf{Frequency separation of strain across the transmission line.} \textbf{a}, Response $\Delta C/C(0)$ versus strain for each sensing region, tested individually to determine the gauge factor via a linear fit. \textbf{b}, The simulated capacitance frequency response $C(\varepsilon)-C(0)$ of the model circuit from Fig.~\ref{fig:system}b, using initial values and gauge factors from \textbf{a} under strains of 10\%--40\%. \textbf{c}, Corresponding frequency sweeps obtained with the LCR meter on the strain sensitive fibre samples. \textbf{d}, The response evaluated using our readout system at four discrete frequencies.} \label{fig:spectrum-comparison}
\end{figure}

Each transmission line segment (or discretized stage) acts as a low-pass RC filter. Lower frequency excitation signals have less attenuation through the line, allowing the impedance of the entire line to be measured. Conversely, higher frequency excitation signals are attenuated to a greater degree and predominantly measure the impedance of the sensing area nearer to the readout circuit. Therefore, by comparing the impedance measured at a higher frequency with a lower one, the spatial distribution of capacitance through the line may be inferred. When multiple regions are strained simultaneously, it is plausible that the capacitance change at one region may mask another, resulting in a strain reconstruction that is not unique. Other works have shown bench experiments with successful measurements from all combinations of simultaneously strained regions \cite{Tairych2018,Tairych2017}. We add confidence to these findings from a theoretical perspective by analyzing the structural identifiability of the lumped model from Fig.~\ref{fig:system}b. Structural identifiability is a general mathematical technique to determine if it is possible to uniquely reconstruct the parameters of a system given its input signal and output measurements and is commonly applied to biological compartmental models \cite{Bellman1970} and in structural dynamics \cite{Chatzis2015}. Our analysis (fully detailed in Section~S3 of the Supplementary Materials) indicates that the $R_i$ and $C_i$ parameters of the lumped sensor systems are uniquely measurable using a time-varying input signal. This theoretical method is subject to certain conditions, such as an assumption of perfect noise-free measurements, constant RC parameters, and the use of the lumped model circuit rather than the true distributed parameter system. In reality, there is always noise present and the parameters, although slowly varying compared to the excitation, are not constant. Therefore, the unique identification of the real system's parameters may not be guaranteed. However, our mathematical analysis supports the empirical evidence in this paper and others \cite{Tairych2017} that the localized strain is uniquely reconstructable. Stretchable conductive materials like the conductive polymer coating used here generally have significantly higher resistance compared to typical rigid/metal conductors. We may use the relatively higher resistance to more easily access strain measurements localized in space along the stretchable fibre length, because the larger resistance shifts the frequency response to lower bands. Sensor characterization tests show that the piezoresistance, or increase in resistance versus strain, of the HACS fibre remains low (Fig.~S6c). This is important to ensure that strained segments maintain sufficient conductance so that the subsequent segments further down the fibre may still be measured. We must first choose excitation frequencies---or equivalently, manipulate the sensor's specific resistance and capacitance---to maximize the discrimination between strain across each sensing region. We began an initial test by fabricating four \SI{10}{\centi\metre} long HACS with pitch of approximately \SI{4}{\milli\metre}; this should yield $\mathrm{GF} \approx 0.5$ according to the model and results from our previous study \cite{Cuthbert2023}. We measured the sensor response $\frac{\Delta C}{C(0)}$ versus strain for each sensor independently and fit a linear model to obtain relaxed capacitance $C(0)$, relaxed resistance $R(0)$, and GF for each (Fig.~\ref{fig:spectrum-comparison}a). We then modelled the impedance of the cascaded transmission line connection of the sensors from Fig.~\ref{fig:system}a at arbitrary strain using the Cauer RC ladder transfer function \cite{Hwang1984} from equation~(\ref{eq:transmission-line}). We allowed the resistance and capacitance parameters to vary with strain, assuming a linear relationship obtained from the single sensor tests, for example: 

\begin{equation*}
	C(\varepsilon) = C(0)\left(1 + \varepsilon \cdot GF\right).
\end{equation*}

The frequency response of the model at strains of 10\%--40\% are shown in Fig.~\ref{fig:spectrum-comparison}b, presented as the change in RC-parallel capacitance compared to that at $\varepsilon=0$. We observed that the frequency window \SI{10}{\kilo\hertz}--\SI{100}{\kilo\hertz} differentiates sensor segments II--IV well with respect to strain. At frequencies below this window, the observed $\Delta C$ versus strain is large, but nearly identical behaviour is seen at each segment I--IV. Above this window, the $\Delta C$ versus strain is uniformly low for all sensor segments. We then connected the individual segments in cascaded fashion (following the schematic of Fig.~\ref{fig:system}b) and placed them in a fixture to confirm our model predictions. The fixture allows each sensing region to be strained independently. For the purposes of illustrating the mechanism and for device validation, we strain only one region at a time. In Section~\ref{sec:localized-strain} \emph{Localized Strain Reconstruction}, we will also examine the case where two regions are strained simultaneously. The individual sensor segments were connected with low-resistance copper wire for during testing so that the exact same samples tested individually could also be tested in series. This is electrically equivalent to a continuous strain sensing fibre, which we used in the prototype garment that follows. An LCR meter was used to perform frequency sweeps of the system to obtain the data shown in Fig.~\ref{fig:spectrum-comparison}c, which agree well with our simple model. As expected, as we continue down the ladder network from the readout circuit (from segment I to IV), the change in capacitance response attributable to strain at higher frequencies becomes lower. Finally, we performed a similar measurement using our readout system instead of the LCR. Our device (Fig.~\ref{fig:system}c) functions as a compact, configurable impedance analyzer that measures at multiple frequencies in parallel at high speeds (around \SI{30}{\hertz} output data rate), compared to the LCR that measures across many frequencies with low resolution in time (seconds per sweep). We chose four excitation frequencies (approximately \SI{12.5}{\kilo\hertz}, \SI{25}{\kilo\hertz}, \SI{50}{\kilo\hertz}, and \SI{100}{\kilo\hertz}) from the frequency window we previously identified where the change in capacitance versus strain between each sensor segment is approximately maximal (see Fig.~\ref{fig:spectrum-comparison}b). We observe a similar trend using our device as the LCR sweeps (Fig.~\ref{fig:spectrum-comparison}d). We have now established a basis for how distributed sensing may be accomplished using impedance measurements at multiple frequencies. Next, we continue by briefly describing the signal processing background about how the system measures the impedance at multiple frequencies in parallel and with high readout speeds.

\begin{figure}
	\centering
	\includegraphics[width=88.9mm]{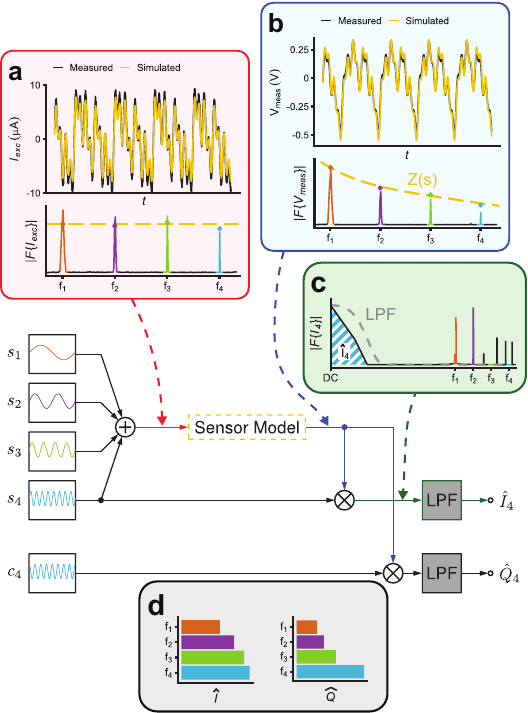}
	\caption{\textbf{Abbreviated schematic with example signals.} \textbf{a}, Top: multi-sinusoid excitation current $i_{exc}(t)$ waveform (simulated and measured) and its Fourier transform. \textbf{b}, Voltage response from the sensor $v_{meas}(t)$ (measured, and simulated using the ladder model from equation~(\ref{eq:transmission-line})) and its Fourier transform, with sensor model frequency magnitude response overlaid in dashed yellow. \textbf{c}, Fourier transform of the PSD mixer output $I_4[k]$, with DC component $\hat{I}_4[k]$ shown as hatched area and approximate low-pass filter response shown in dashed grey. \textbf{d}, The relative in-phase $\hat{I}_i[k]$ and quadrature $\hat{Q}_i[k]$ impedance components from the sensor model simulation.} \label{fig:signal-flow}
\end{figure}

\paragraph{High-Speed Impedance Measurement} %\label{sec:signal-processing}
Here we describe the impedance measurement procedure used in our readout electronics. The signal processing pathway is outlined in Fig.~\ref{fig:system}d and described below. Fig.~\ref{fig:signal-flow} shows a more simplified system schematic with example signals from both simulation and measurements. A detailed description of each component's implementation in firmware or hardware may be found in Section~S1 of the Supplementary Materials. Additionally, the system calibration process is outlined in Section~S2.

The circuit first includes a sine wave generator (SWG) that produces digital sine and cosine signals at each $f_i$ of $N_f$ frequencies. Using $k$ as a discrete-time index variable and $f_s$ for the sampling frequency, the SWG produces the signals shown in equation~(\ref{eq:sin-cos}) (we use square bracket notation to indicate discrete-time signals and parentheses to indicate continuous-time).

\begin{flalign} \label{eq:sin-cos}
\begin{aligned}
s_i[k] &= \sin{\left(2\pi \frac{f_i}{f_s} k\right)} \quad i \in \{ 1, \ldots, N_f \} \\
c_i[k] &= \cos{\left(2\pi \frac{f_i}{f_s} k\right)} \quad i \in \{ 1, \ldots, N_f \}
\end{aligned}
\end{flalign}

The digital-to-analog converter (DAC) converts the sum of the sine waves to an analog excitation voltage signal $v_{exc}(t)$. This signal satisfies the conditions of our structural identifiability analysis in order to uniquely determine the lumped system's RC parameters. The signal is conditioned and converted to a current signal $i_{exc}(t)$ by the transconductance amplifier (TCA), following equation~(\ref{eq:current-signal}). An example of this current signal is shown in Fig~\ref{fig:signal-flow}a along with its spectrum, consisting of $N_f=4$ frequency components of approximately equal magnitude. For simplicity, we assume a delay-less ADC transfer function equal to unity and approximate the TCA transfer function as a frequency-independent gain $G$. The dynamics introduced by the ADC and TCA are discussed in Supplementary Materials Sections~S1.1--S1.3.

\begin{equation} \label{eq:current-signal}
i_{exc}(t) = G\sum_{i=1}^{N_f} s_i(t) = G\sum_{i=1}^{N_f} \sin{(2\pi f_i t)}
\end{equation}

The linear, time-invariant impedance response of the sensor $Z(s)$ from equation~(\ref{eq:transmission-line}) imparts a frequency-dependent gain and phase shift for each component of the $i_{exc}(t)$ excitation signal as shown in equation~(\ref{eq:post-sensor}), where $|Z_i|$ is the gain and $\varphi_i$ is the phase shift at frequency $f_i$. 

\begin{flalign} \label{eq:post-sensor}
\begin{aligned}
v_{meas}(t) &= \mathcal{L}^{-1}\{ Z(s) I_{exc}(s) \} \\
&= G \sum_{i=1}^{N_f} |Z_i| \sin{(2\pi f_i t + \varphi_i)}
\end{aligned}
\end{flalign}

In Fig.~\ref{fig:signal-flow}b, the resulting voltage across the sensor $v_{meas}(t)$ is shown. The magnitudes of the four frequency components are modified following the linear response of the sensor. This signal is digitized and processed by a phase-sensitive demodulator (PSD) which uses the pure sine and cosine waves from equation~(\ref{eq:sin-cos}) as a reference to extract in-phase and quadrature components at each frequency. Using Euler's formula, the signals from equation~(\ref{eq:sin-cos}) may be written as:

\begin{align}
s_i[k] &= \frac{e^{j2\pi \frac{f_i}{f_s} k}-e^{-j2\pi \frac{f_i}{f_s} k}}{2j} \label{eq:euler-sin} \\
c_i[k] &= \frac{e^{j2\pi \frac{f_i}{f_s} k}+e^{-j2\pi \frac{f_i}{f_s} k}}{2}, \label{eq:euler-cos}
\end{align}

and each frequency term of equation~(\ref{eq:post-sensor}) may be written as shown in equation~(\ref{eq:euler-vmeas}).

\begin{equation} \label{eq:euler-vmeas}
v_{meas}[k] = G \sum_{i=1}^{N_f} |Z_i| \frac{e^{j2\pi \frac{f_i}{f_s} k + \varphi}-e^{-j2\pi \frac{f_i}{f_s} k - \varphi_i}}{2j}
\end{equation}

The PSD mixer multiplies each term $i$ of equation~(\ref{eq:euler-sin}) with equation~(\ref{eq:euler-vmeas}) to calculate the in-phase component. After simplification, we obtain equation~(\ref{eq:inphase}).

\begin{align}
I_i[k] &= s_i[k] \cdot v_{meas}[k] \nonumber \\
&= \frac{G|Z_i|}{2} \left( -\frac{e^{j2\pi \frac{2f_i}{f_s} k + \varphi_i} + e^{-j2\pi \frac{2f_i}{f_s} k + \varphi_i}}{2} + \frac{e^{\varphi_i} + e^{-\varphi_i}}{2} \right) \nonumber \\
&= -\frac{G|Z_i|}{2}\cos{\left(2\pi \frac{2f_i}{f_s} k + \varphi_i\right)} + \underbrace{\frac{G\cdot |Z_i|}{2}\cos{(\varphi_i)}}_{\hat{I}_i[k]} \label{eq:inphase}
\end{align}

Similarly, the PSD mixer multiplies each term $i$ of equation~(\ref{eq:euler-cos}) with equation~(\ref{eq:euler-vmeas}) to yield the quadrature component, equation~(\ref{eq:quadrature}).

\begin{align}
	Q_i[k] &= c_i[k] \cdot v_{meas}[k] \nonumber \\
	&= \frac{G|Z_i|}{2} \left( \frac{e^{j2\pi \frac{2f_i}{f_s} k + \varphi_i} - e^{-j2\pi \frac{2f_i}{f_s} k - \varphi_i}}{2j} + \frac{e^{\varphi_i} - e^{-\varphi_i}}{2j} \right) \nonumber \\
	&= \frac{G|Z_i|}{2}\sin{\left(2\pi \frac{2f_i}{f_s} k + \varphi_i\right)} + \underbrace{\frac{G\cdot |Z_i|}{2}\sin{(\varphi_i)}}_{\hat{Q}_i[k]} \label{eq:quadrature}
\end{align}

Naturally, equations (\ref{eq:inphase}) and (\ref{eq:quadrature}) contain additional product terms for interactions between all $N_f$ frequency components, but these are omitted for clarity as they will be filtered out in the following step. The mixing process is exemplified in Fig.~\ref{fig:signal-flow}c for the in-phase impedance corresponding to $f_4$: $I_4[k]$. The spectrum of the output of the mixer contains the baseband, time independent term of equations (\ref{eq:inphase}) and (\ref{eq:quadrature}) as well as the high-frequency products. Let $\hat{I}_i[k]$ and $\hat{Q}_i[k]$ be the DC components extracted through an ideal low-pass filter. The real filter applied to the signals is discussed in Section~S1.5 of the Supplementary Materials. A communications block outputs $\hat{I}_i[k]$ and $\hat{Q}_i[k]$ to the client software running on a PC. The result (Fig.~\ref{fig:signal-flow}d) may be converted to magnitude and phase quantities using equation~(\ref{eq:mag-phase}), or to any passive 1-port circuit model (e.g. RC parallel). In the following section, we explain our method to obtain localized strain from the $n$-frequency impedance signals $\hat{I}$, $\hat{Q}$.

\begin{flalign}
\begin{aligned} \label{eq:mag-phase}
|Z_i| &= \frac{2}{G}\sqrt{\hat{I}_i^2[k] + \hat{Q}_i^2[k]} \\
\varphi_i &= \tan^{-1}{\left(\frac{\hat{Q}_i[k]}{\hat{I}_i[k]}\right)}
\end{aligned}
\end{flalign}

\paragraph{Localized Strain Reconstruction} \label{sec:localized-strain}
Several methods have been used to reconstruct capacitance (and thus strain) from the electrical impedance data shown in equation~(\ref{eq:mag-phase}). A graphical approach completed by solving a linear system of equations arising from the change of capacitance at the excitation frequencies is fast and efficient \cite{Tairych2017}, but works best when the resistance of each segment remains constant. With most types of stretchable capacitive sensors, the compliant electrodes have strain-dependent piezoresitive behaviour causing this assumption to be violated. Alternatively, optimization methods may solve for the unknown capacitances (and even resistances) of the model's system of nonlinear equations \cite{Tairych2018}. In this case, the input resistance $\mathfrak{R}\left\{Z_1(\omega)\right\}$ and reactance $\mathfrak{I}\left\{Z_1(\omega)\right\}$ of the transmission line at each of $N_f$ frequencies are measured, following equation (\ref{eq:transmission-line}). From these $2\cdot N_f$ measurements, $N_f$ unknown resistances and $N_f$ unknown capacitances are estimated. Conventional structured system identification algorithms such as subspace identification may be used to identify the capacitance values in real-time \cite{Calzavara2021}. These methods require a non-convex optimization problem to be solved at every time step, which may be impractical for high-speed online measurement. The optimization is not guaranteed to converge and often converges to a local optimum as the number of identified parameters increases \cite{Ljung2019}. Other approaches have used machine learning methods such as support vector machines or artificial neural networks \cite{Sonar2018, White2017}. These techniques are powerful but rely on the extensive collection of training data. Once trained, they are not easily modifiable to use with different sensors or number of segments.

We chose to use machine learning, specifically a multilayer perceptron (MLP) algorithm, to reconstruct strain at each sensor segment from the impedance signals. Despite the drawbacks mentioned above, MLP was selected because the sensors have non-linear response and thus one model could both produce strain estimates and compensate for sensor non-idealities. Often strain is not the outcome of interest and is merely a proxy for another signal (e.g., joint angle), which is then predicted using a machine learning algorithm. The MLP may operate directly on the impedance signals to compute the final measurement without using strain or capacitance as an intermediate step. Neural network-based models have shown good results in previous work for refining strain sensor signals \cite{Rezaei2019, Hannigan2021}. The MLP model avoids the need for online optimization and once trained, inference can run very quickly. Our MLP architecture incorporates $\ell_1$ and $\ell_2$ norm weighted regularization penalties. While tuned empirically, the former is intended to encourage sparsity in the output to reduce false positive strain signals when the segment is indeed at rest. The latter is applied as a general regularizer to reduce the degree of overfitting. Further details about the data collection protocol and model architecture are shown in the Materials and Methods section.

%\section{System Validation Experiment} \label{sec:validation} % Decided to remove this heading as it's related to the previous section quite well.

\begin{figure}
	\centering
	\includegraphics[width=88.9mm]{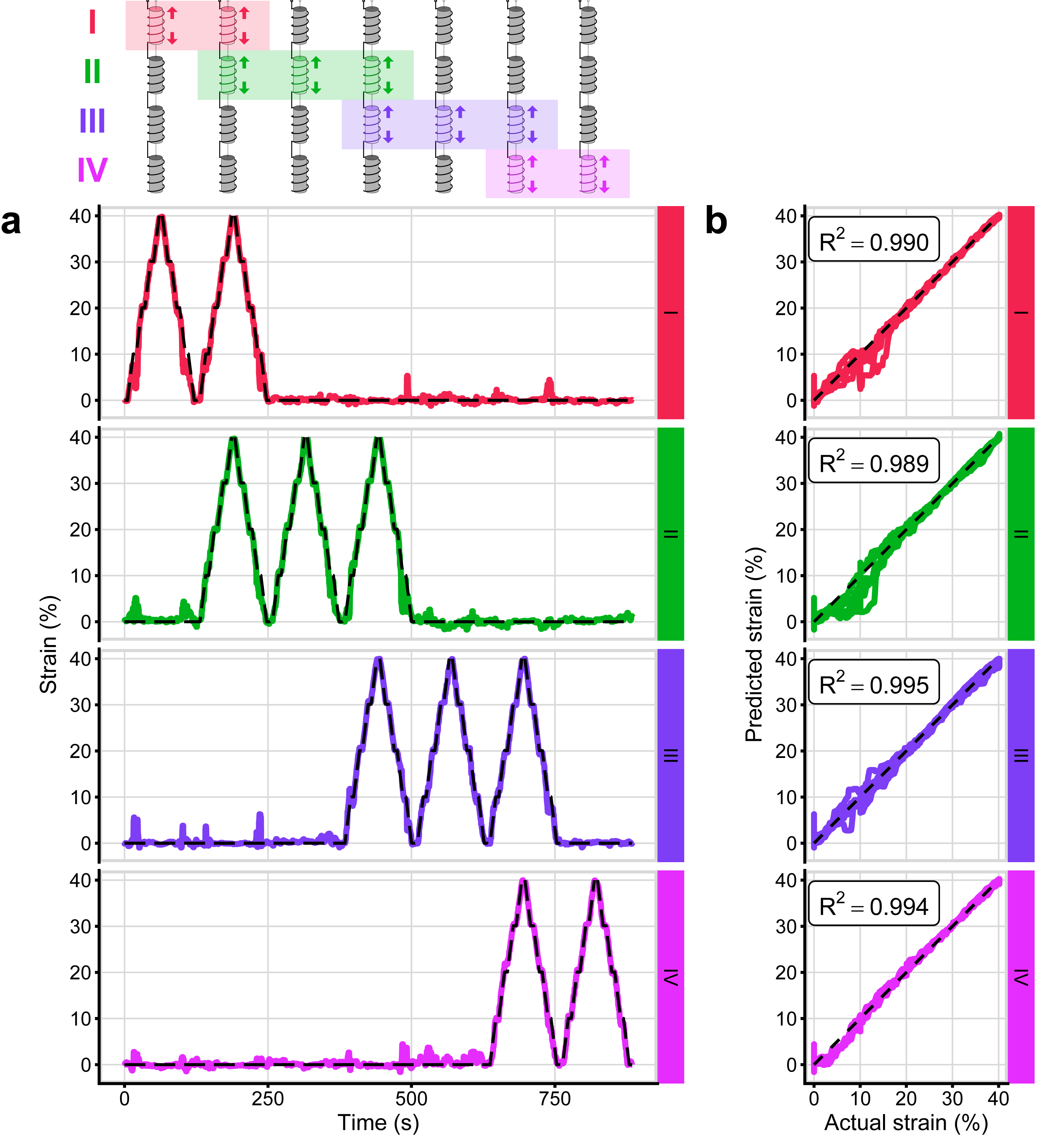} % Using PNG image because EPS file is too large.
	\caption{\textbf{Strain reconstruction tests.} \textbf{a}, Reconstructed (solid) and UTM reference (dashed black) strain for each sensor segment I--IV during the system validation experiment (test set shown); \textbf{b}, Scatterplots of correlation between predicted and reference strain for each sensor, and their associated R\textsuperscript{2} values.} \label{fig:strain-reconstruction}
\end{figure}

Having designed a system capable of high-speed impedance measurement to enable distributed sensing, we ran a bench experiment to validate the device, collect training data, and quantify strain reconstruction accuracy. We placed the four HACS fibres described previously in the test fixture and subjected them to various combinations of strains (see the Materials and Methods section and Section~S6 of the Supplementary Materials for test procedure and experimental set-up details, respectively). We trained a simple MLP model (full details in the Materials and Methods section) to directly estimate the strain localized to each of the four segments from the in-phase and quadrature impedance components at each of the four excitation frequencies $\hat{I}_i[k], \hat{Q}_i[k] \quad i \in \{1, \ldots, 4\}$ (totally, 8 inputs and 4 outputs). We found a test set strain reconstruction root-mean-squared error (RMSE) of 1.04\% with coefficient of determination (R\textsuperscript{2}) of 0.992. The test set actual and predicted strain patterns are shown in Fig.~\ref{fig:strain-reconstruction} and full results are listed in Table~S4. We noticed that the accuracy was generally higher on segments located more distal from the reader along the chain. This result may be explained by the technical specifics of the readout circuit. The lower-frequency excitation signal that preferentially measures the distal segment is more highly oversampled, increasing signal-to-noise ratio (SNR; see Section~S1 of the Supplementary Materials for details). We also compared the reconstruction accuracy between the cases with one and two segments were simultaneously strained. In the training and validation sets, reconstruction RMSE was approximately 30\% higher when straining two segments simultaneously, but this was not seen in the test set. This behaviour possibly indicates the model over-fitting to single segment straining, which was more frequent in the data set. From the correlation plots in Fig.~\ref{fig:strain-reconstruction}b, it is evident that goodness of fit increases with strain in all four segments. This observation is partially attributable to variability in the segments' length at the rest position, such that they do not all begin at exactly 0\% strain.

\begin{figure}
	\makebox[\textwidth][c]{\includegraphics[width=180mm]{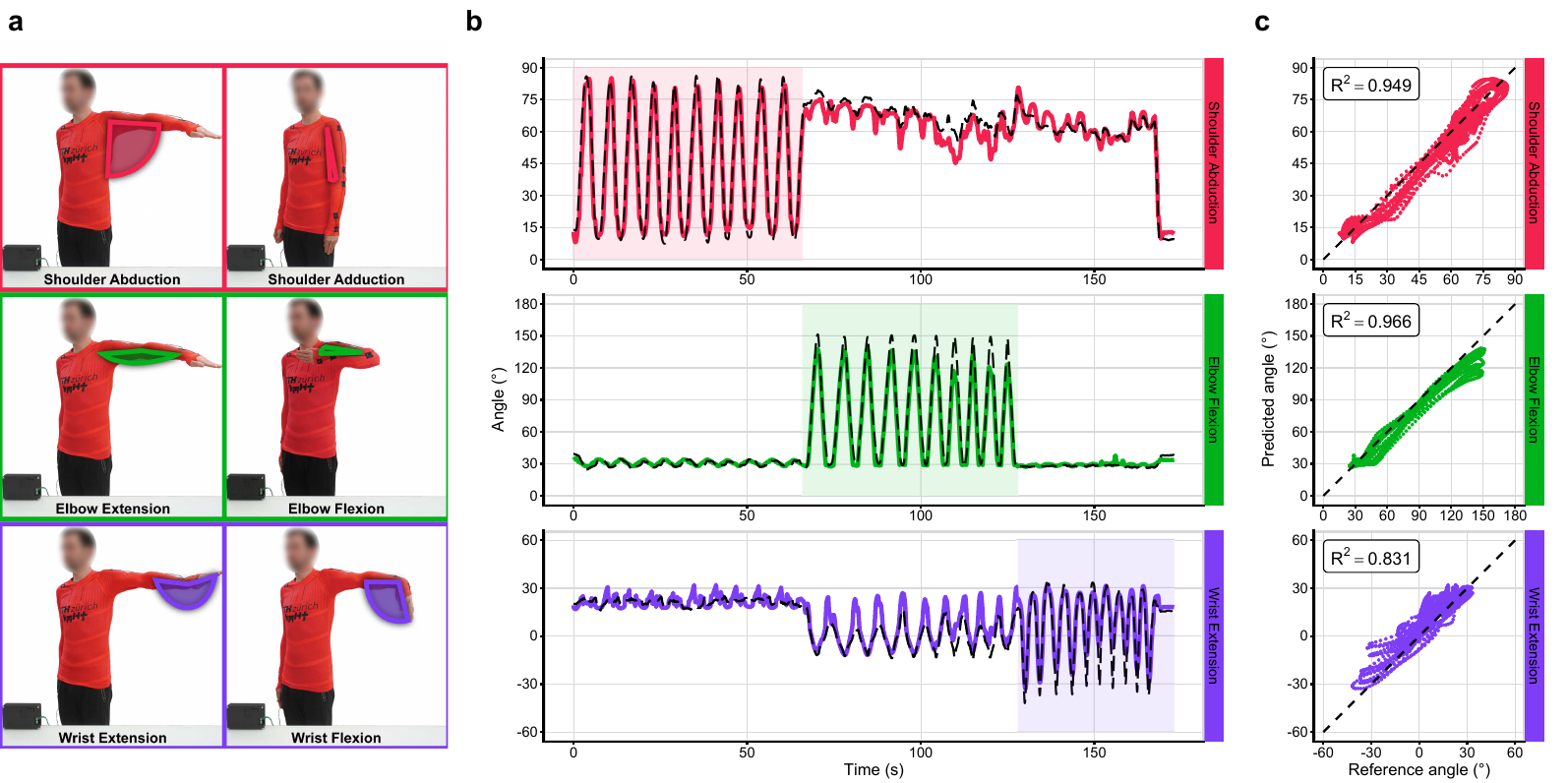}}
	\caption{\textbf{Joint angle monitoring with the wearable prototype.} \textbf{a}, Visual demonstration of the three joint angles under consideration. \textbf{b}, Predicted (solid) and optical motion capture reference (dashed black) angles for each joint, from the final fold of testing, where shaded regions indicate the part of the test where that joint was voluntarily moved. \textbf{c}, Scatter plots of predicted versus reference angles with associated R\textsuperscript{2} values, where the black dashed line indicates perfect correlation.} \label{fig:angle-reconstruction}
\end{figure}

\paragraph{Joint Angle Monitoring} \label{sec:joint-angle-monitoring}
After observing promising results from the validation test under controlled conditions with 4 segments, we began a test of the previously introduced instrumented garment to reconstruct arm joint angles from the single sensing fibre. The participant performed three movements: shoulder adduction/abduction from approximately \SI{0}{\degree}--\SI{90}{\degree} with respect to the torso, elbow flexion throughout its entire range (approximately \SI{180}{\degree}), and wrist flexion from approximately \SI{0}{\degree}--\SI{45}{\degree} with respect to the forearm axis. These movements were chosen to capture the three major arm joints with a relatively large range of motion. Although the sensing fibre has been shown to have sufficient frequency response to detect strain rates above \SI{5}{\hertz} \cite{Cuthbert2023}, we attempted to further limit the frequency of the movements to below \SI{1}{\hertz} to avoid rate-dependent effects. Reference joint angles were measured using a gold standard infrared camera optical motion capture (OMC) system. %While the subject attempted to keep each movement relatively isolated, there is still considerable crossover of one joint movement to the others (in both our system's measurement and the reference). 
We trained a similar MLP model to that previously described, this time for joint angle prediction rather than strain (see the Materials and Methods section for details). We show in Fig.~\ref{fig:angle-reconstruction} the results from the last set of test data. We achieved joint angle regression RMSEs of \SI{4.9}{\degree}, \SI{6.5}{\degree}, and \SI{6.1}{\degree} for the shoulder, elbow, and wrist, respectively. As a percentage of the approximate ranges of motion, the normalized RMSE was 6.5\%, 5.4\%, and 8.1\%. R\textsuperscript{2} values were found to be 0.949, 0.966, and 0.831, with full cross-validation plots and results available in Fig.~S11 and Table~S5. In this work, limited data from one participant and simple data processing methods were used to show the concept of distributed sensing in a wearable application without placing too much focus on maximizing joint angle accuracies. Nevertheless, our single fibre system approaches the RMSE values of around \SI{2}{\degree}--\SI{5}{\degree} reported for joint angle monitoring with multiple conventional, redundant sensors clustered around the joint \cite{Esfahani2018,Jin2020} (albeit, under cases of well-controlled movements), yet without the associated wiring and connection hurdles. It should be noted that the OMC markers were placed on top of the instrumented shirt. Although tight-fitting, we expect errors from the motion capture measurements because of marker displacements to be higher than standard biomechanics protocols with markers placed on the skin. The distributed sensing technology is naturally extensible to having more sensing regions per joint in a dense area without additional discrete sensors nor connections. We see that the reconstructed joint angles sometimes fail to track the peaks of the reference (see, e.g., the shoulder traces from Fig.~S11). When this occurs during extension peaks, it is probably that some of the sensors had insufficient pre-strain to avoid becoming slack at full extension. The wrist had the poorest tracking accuracy, but also was the most difficult to accurately record the OMC reference because the entire hand is resolved as a rigid body with only one marker. Our prototype garment is the first wearable device that uses distributed strain sensing to estimate multiple joint angles using a single fibre sensor. We believe that with more extensive and accurate data collection, the technology has great promise to allow fibres in clothing to be used as sensors in a fully textile, unobtrusive package.

\section{Discussion}
In this work, we have reported a distributed sensing technology and demonstrated its use in a wearable application. We combined an electronic system with our capacitive fibre strain sensor to enable distributed sensing in textiles. We have developed the first compact impedance analyzer circuit custom designed for distributed strain sensing in a wearable context. We apply the sensors in a novel way that allows sensitivity to be modulated across the fibre length to reject unwanted strain signals, such as those arising at undesired areas or from folds in the garment. The localized sensitivity of the desired sites allows a minimal number of frequencies and simpler electronics to reconstruct strain with around 1\% error in controlled bench tests. Our device is validated with a UTM-controlled fixture to present quantitative strain reconstruction results, whereas previous distributed strain sensing reports have mostly shown qualitative tests. Applied in a prototype garment, our system estimates multiple joint angles with a single fibre, often to within \SI{5}{\degree} of an OMC system for major joint axes of interest. 

As this is primarily a study to first demonstrate distributed sensing in a wearable device, there are certain limitations that may be addressed in future research. Scaling up the number of sensor segments is an obvious evolution of the technology. Though shown with one fibre and three joint measurements, the distributed sensing technique is naturally extensible to multiple fibres each having many measurement sites. This will be of focus for future research, exploring the limit of sensing region density. One can envision the future development of multi-sensing fabrics capable of measuring strain maps across the body in real time. There remain questions about the effect on strain reconstruction accuracy when multiple sensing regions are strained at once under non-ideal conditions. We observed a moderate decrease in test set accuracy in the case of simultaneous strain, but we only explored the case of straining two adjacent segments. Evidence in other studies as well as our structural identifiability analysis supports the claim that the masking of one sensor segment by another appears not to be a major obstacle to the practical use of this distributed strain sensing system, but further investigation would be of high interest. Although the MLP model we used to reconstruct strain and joint angles from impedance data is fast to run during inference, able to correct for nonlinear effects of the sensors, and able to compute the desired outcome measure directly, using supervised machine learning algorithms for this task introduces limitations. In our implementation, the algorithm must be re-trained for use with different sensors types or number and positions of segments, as it has no innate knowledge of the dynamics. Training requires pre-collected data. Thus, one loses dynamic reconfigurability with this model, which is an important benefit of distributed sensing and of great interest for future work. Future research may target improved algorithms for capacitance reconstruction such that the size and number of segments can be dynamically changed. Finally, we anticipate a further study including testing to evaluate the robustness of distributed sensing in more real-world scenarios, for example, with tracking of random or everyday arm movements.

While we aimed to keep the electronics simple by using an FPGA and minimal analog circuitry to allow translation to a fully on-chip solution more easily; it would be relatively straightforward to implement a wireless battery-powered system in the future. The electronics as designed could be made much smaller to become truly portable by integrating everything on a single PCB, removing extraneous components, and changing the form factor so that it better fits a compact enclosure. Despite this, challenges exist with miniaturization of an accurate, wide-band impedance spectroscopy device. A main benefit to this conceptualization of distributed textile sensing is that the complex electronics may be decoupled from the all-textile garment, interfaced with only two connection points. The elimination of connection points within the garment and localization of all remaining connections in a proximal hub drastically simplifies e-textile production, facilitating the progress of all-textile wearable sensor systems.

\section{Materials and Methods} \label{sec:methods}

\paragraph{Fibre Sensor and Prototype Garment Fabrication} %\label{sec:methods-sensor-fibre}
The helical auxetic capacitive sensors were fabricated as described in \cite{Cuthbert2023}. A stretchable conductive fibre (with diameter \SI{0.75}{\milli\metre}), produced by applying a polypyrrole coating to an elastic fibre, was wound with a coil of 35~AWG insulated copper wire. The core functions as one electrode of a capacitor with relatively high resistance and the coil functions as the other electrode with low resistance. The enamel insulation on the copper wire provides a dielectric layer. A thicker core was formed with three parallel coated elastic fibres for a final diameter of ca. \SI{2}{\milli\metre}. For the bench tests, four separate core segments of relaxed length $\ell_0=\SI{10}{\centi\metre}$ were wound with \SI{4}{\milli\metre} pitch. For the prototype garment, a continuous core of \SI{80}{\centi\metre} was wound with three sensitive regions of relaxed length $\ell_0=\SI{10}{\centi\metre}$ and \SI{8}{\milli\metre} pitch. The sensing fibre was anchored with approximately 10\% pre-strain when the joint of interest was extended. Between the sensitive regions were two insensitive regions of varying lengths and a \SI{3.5}{\milli\metre} pitch. The continuous sensor had an unstrained linear DC core resistance of \SI{0.75}{\kilo\ohm\per\centi\metre} and capacitance of \SI{4.7}{\pico\farad\per\centi\metre}. Capacitance and resistance versus strain of a sample sensor as measured with the LCR and our device is presented in Fig.~S6. For integration in the garment, we spaced the sensitive regions empirically so that they were centered around and fully overlapped the shoulder, elbow, and wrist joints. The sensing fibre was attached to the garment using clips sewn to elastane fabric backing patches. Photographs of the donned garment may be seen in Figure~S10 of the Supplementary Materials.

\paragraph{Electronics Design and Implementation}
The digital signal processing (SWG, DAC, and PSD) was prototyped in Simulink (The Mathworks, Natick, MA, USA) and implemented with custom Verilog code on field-programmable gate array (FPGA) fabric (Lattice iCE40HX-8K; Lattice Semiconductor, Hillsboro, OR, USA). The analog circuitry (TCA, ADC) was prototyped in SPICE and was realized on a custom \SI{75}{\milli\metre} $\times$ \SI{45}{\milli\metre} PCB shown in Fig.~\ref{fig:system}c. Further details are provided in Section~S1 of the Supplementary Materials. Verilog code, board layout files, client software, SPICE simulations, and mechanical parts files used in this project are available in our repository: \url{https://gitlab.ethz.ch/BMHT/textile-wearables/textile-electronics/sensor-readout-board}.

\paragraph{Validation Data Collection Protocol} %\label{sec:methods-validation-collection}
A universal testing machine (UTM) (Instron E3000; Instron, Norwood, MA, USA) was interfaced with a custom-made fixture that allows straining of each sensing region independently (see Fig.~S9). Four fibre strain sensor segments I--IV were mounted on sliding rails with a centre section that is connected to the UTM crosshead. The entire fixture top plate was then manually adjusted to bring each sensing region to approximately zero strain. One or two adjacent sensing regions were positioned in the centre section of the fixture and thus subject to strain at a given time (allowing simultaneous combinations of segments I-II, II-III, and III-IV). The readout electronics are synchronized to the UTM with a electronic trigger signal and the data is recorded by the client software running on a PC. The sensor was characterized electrically using an inductance-capacitance-resistance (LCR) meter as a reference (Hioki IM3536; Hioki, Ueda, Nagano, Japan). The UTM was programmed to follow an up-down staircase strain pattern, with steps of 10\%, a ramp rate of \SI{1}{\percent\per\second}, and a maximum strain of 40\%. At each step, the strain was held for \SI{5}{\second}. Because of limitations of the fixure, we could only perform the simultaneous strain combinations A-B, B-C, and C-D. The staircase pattern was repeated 6 times per sensing region combination. Data and scripts required to reproduce the figures in this manuscript are available in our repository: \url{https://gitlab.ethz.ch/BMHT/publications/distributed-sensing-along-fibres}.

\paragraph{Strain Reconstruction Model} %\label{sec:methods-strain-model}
A very simple multi-layer perceptron (MLP) model was compiled using Keras/Tensorflow with architecture consisting of an 8-unit input layer, two hidden layers of sizes (16, 32) units and using tanh and ReLU activation, respectively, and a 4-unit output layer with linear activation. The second hidden layer had $\ell_2$ regularization with weight $1\times 10^{-5}$. The six trials collected as described in the above section were split into training (trials 1--4), validation (trial 5), and test (trial 6) sets. The model was trained with batch size 256 using the Adagrad optimizer with learning rate 0.1. Training was terminated when 50 epochs elapsed without an improvement in validation set mean squared error (MSE). The predicted strain model output was filtered with a moving median filter (window size \SI{2}{\second}) prior to computing the metrics presented in Table~S4.

\paragraph{Wearable Prototype Data Collection Protocol} %\label{sec:methods-prototype-collection}
The subject donned the upper-body garment and reflective optical motion capture (OMC, Vicon Ltd., Oxford, UK) markers were placed according to the upper-body Plug-In Gait model. The subject performed 10 repetitions each of shoulder adduction, elbow flexion, and wrist flexion. These movements were repeated for 10 sets, with a brief rest period in between. The impedance signals from the readout board were synchronized electronically with the OMC system using its trigger output. The joint angle reference signals were pre-processed with a Savitzky-Golay filter (\SI{2}{\second} window length, 4\textsuperscript{th}-order polynomial) and the impedance signals were Butterworth low-pass filtered (4\textsuperscript{th}-order, \SI{2}{\hertz} cutoff).

\paragraph{Joint Angle Reconstruction Model} %\label{sec:methods-angle-model}
The MLP model used for strain reconstruction was slightly modified to be used in joint angle reconstruction. The output layer was adjusted to three units, corresponding to the three joint angles of interest. $\ell_1$ and $\ell_2$ regularization was applied to the second hidden layer with weights 0.04 and 0.15, respectively, and both hidden layers used ReLU activations. The 10 trials of data collected as described in the above paragraph underwent a leave-one-out cross-validation scheme. At each cross-validation round, 1 trial was reserved for test data and the remaining 9 were split into test (80\%) and validation (20\%) sets. The input data were Z-score normalized (with scaling determined from training set only). In addition, the 10\textsuperscript{th} epoch was reserved for testing; the results of this set are presented in the text and no model tuning was performed using test data. The model was trained with batch size 64 using the Adam optimizer with learning rate 0.005. Training was terminated when a weighted average (shoulder: 1/3, elbow: 1/3, wrist: 2/3) of validation set MSE  did not improve for 250 consecutive epochs.

% Your references go at the end of the main text, and before the
% figures.  For this document we've used BibTeX, the .bib file
% scibib.bib, and the .bst file Science.bst.  The package scicite.sty
% was included to format the reference numbers according to *Science*
% style.

%BibTeX users: After compilation, comment out the following two lines and paste in
% the generated .bbl file. 

%\bibliography{}
%\bibliographystyle{Science}

\let\oldthebibliography=\thebibliography
\let\oldendthebibliography=\endthebibliography
\renewenvironment{thebibliography}[1]{
	\oldthebibliography{#1}
	\setcounter{enumiv}{43}
}{\oldendthebibliography}
\vspace{-0.62\baselineskip}
{
	\renewcommand{\section}[2]{}%

}

\section*{Acknowledgements}

The authors would like to acknowledge the help of Marina Geissmann at the Swiss Centre for Movement Analysis (SCMA) whose expertise in recording and processing the optical motion capture reference data presented in Section~\ref{sec:joint-angle-monitoring} \emph{Joint Angle Monitoring} was greatly appreciated. 

\subsection*{Funding}

The authors acknowledge that they received no funding in support for this research.

\subsection*{Competing Interests}

Authors T.J.C. and C.M. are co-inventors on a patent that covers the helical auxetic capacitive sensor technology from reference \cite{Cuthbert2023}. The technology from \cite{Cuthbert2023} was used as cited in this manuscript.

\subsection*{Author Contributions}

\textbf{B.C.H.:} Conceptualization, Methodology, Software, Validation, Formal analysis, Investigation, Data curation, Writing - original draft, Writing review \& editing, Visualization. \textbf{T.J.C:} Conceptualization, Methodology, Validation, Investigation, Writing - review \& editing. \textbf{C.A.:} Methodology, Formal analysis, Software, Data curation, Writing - review \& editing. \textbf{C.M.:} Conceptualization, Methodology, Resources, Writing - review \& editing, Supervision, Project administration, Funding acquisition.

\subsection*{Data and Material Availability}

All data needed to evaluate the conclusions in the paper are present in the paper, Supplementary Materials, and repositories listed in the Materials and Methods section. Data and scripts required to reproduce the figures in this manuscript are located in our repository corresponding to this publication: \url{https://gitlab.ethz.ch/BMHT/publications/distributed-sensing-along-fibres}. Software, firmware, and hardware designs for the readout system are located in our additional repository: \url{https://gitlab.ethz.ch/BMHT/textile-wearables/textile-electronics/sensor-readout-board}. Materials used in this study were not subject to restrictive patents nor materials transfer agreements.

%Here you should list the contents of your Supplementary Materials -- below is an example. 
%You should include a list of Supplementary figures, Tables, and any references that appear only in the SM. 
%Note that the reference numbering continues from the main text to the SM.
% In the example below, Refs. 4-10 were cited only in the SM.     
\section*{Supplementary Materials}

Supplementary Text:\\
\indent S1 Circuit Design Details\\
\indent S2 System Calibration Procedure\\
\indent S3 Identifiability Analysis of the RC Ladder System\\
\indent S4 Effect of Bending on Sensor Response\\
\indent S5 Cyclic Durability Test\\
\indent S6 Validation Experiment Set-Up\\
\indent S7 Full Strain Reconstruction Results\\
\indent S8 Joint Angle Monitoring Experiment Set-Up\\
\indent S9 Full Joint Angle Regression Results\\
Figs. S1 to S11\\
Tables S1 to S5\\
References \textit{(44--61)}

% For your review copy (i.e., the file you initially send in for
% evaluation), you can use the {figure} environment and the
% \includegraphics command to stream your figures into the text, placing
% all figures at the end.  For the final, revised manuscript for
% acceptance and production, however, PostScript or other graphics
% should not be streamed into your compliled file.  Instead, set
% captions as simple paragraphs (with a \noindent tag), setting them
% off from the rest of the text with a \clearpage as shown  below, and
% submit figures as separate files according to the Art Department's
% instructions.

\clearpage

\end{document}